\documentclass[prl,twocolumn,showpacs,floatfix,amsmath,nofootinbib,amssymb,floatfix]{revtex4}
\usepackage{graphicx,color,dcolumn,booktabs,bm}
\usepackage{longtable,lscape}
\usepackage{txfonts}
\usepackage{overpic}
\usepackage{amssymb}
\usepackage{indentfirst}
\usepackage{feynmf}   
\usepackage{slashed}  
\usepackage{cases}
\usepackage{color}
\usepackage{multirow}
\usepackage{epstopdf}
\usepackage{graphicx,color,dcolumn,booktabs,bm}
\usepackage[colorlinks, citecolor=blue,anchorcolor=red,menucolor=red, linkcolor=red,filecolor=red,runcolor=red,urlcolor=blue,frenchlinks=red]{hyperref}

\graphicspath{{Figures/}} %

\begin{document}

\title{Unified Fano-like interference picture for charmonium-like states $Y(4008)$, $Y(4260)$ and $Y(4360)$}

\author{Dian-Yong Chen}\email{chendy@impcas.ac.cn}
\affiliation{
Institute of Modern Physics, Chinese Academy of Sciences, Lanzhou 730000, China}
\author{Xiang Liu$^{1,2}$}\email{xiangliu@lzu.edu.cn (corresponding author: )}
\affiliation{
$^1$School of Physical Science and Technology, Lanzhou University,
Lanzhou 730000, China\\
$^2$Research Center for Hadron and CSR Physics, Lanzhou
University and Institute of Modern Physics of CAS,
Lanzhou 730000, China}
\author{Xue-Qian Li}\email{lixq@nankai.edu.cn}
\affiliation{
School of Physics, Nankai University, Tianjin, 300071, China}
\author{Hong-Wei Ke}
\affiliation{School of Science, Tianjin University, Tianjin 300072, China}
\begin{abstract}
We propose that the unified Fano-like interference picture applies to $e^+e^-\to \pi^+\pi^- J/\psi$ and $e^+e^-\to \pi^+\pi^- \psi(3686)$, where $Y(4260)$ and $Y(4360)$ are observed, respectively, to provide a reasonable interpretation of the asymmetric lines hapes of $Y(4260)$ and $Y(4360)$ structures. Moreover, the Fano-like interference induces an extra broad structure $Y(4008)$ in $e^+e^-\to \pi^+\pi^- J/\psi$ as a companion peak to $Y(4260)$. Three charmonium-like states $Y(4008)$, $Y(4260)$ and $Y(4360)$ observed in $e^+e^-$ annihilation processes are not genuine resonances. Under this scenario, it is well explained why $Y(4008)$, $Y(4260)$ and $Y(4360)$ are absent in the experimental data of the $R$ value scan and missing in open-charm decay channels.
Although the present work presents a typical application of the Fano-like interference in particle physics at the lower energy region, we have reason to believe that the Fano-like phenomena may exist in other processes such as $e^+e^-\to \pi^+\pi^-\psi(3770)$ and $e^+e^-\to K^+K^- J/\psi$, which will be accessible at BESIII and forthcoming BelleII.

\end{abstract}

 \maketitle

{\it INTRODUCTION.---}In the past 12 years, dozens of charmonium-like states were observed by several major particle physics experiments such as the CLEOc, BaBar, Belle, CDF, D$\O$, LHCb, BESIII and so on (see Ref. \cite{Liu:2013waa} for a concise review). With the experimental progress, the family of charmonium-like states has more and more expanded. These observations have been continuing to surprise us with novel discoveries of the exotic properties, and inspire theorists' extensive interest in exploring the underlying mechanism behind those peculiar phenomena. Because these observations of charmonium-like states are closely related to non-perturbative effects of quantum chromodynamics (QCD), the study on charmonium-like states provides us a good chance to deepen our understanding of the QCD confinement, which is one of the most important issues in particle physics.

As one of the most famous charmonium-like states, $Y(4260)$ observed right after $X(3872)$, was reported by the BaBar Collaboration in 2004 via $e^+e^-\to \pi^+\pi^- J/\psi$ \cite{Aubert:2005rm}. Later, the Belle Collaboration confirmed it \cite{Yuan:2007sj, Liu:2013dau}, and indicated that at its vicinity there exists a broad structure $Y(4008)$,  which has mass $M=4008\pm40^{+11}_{-28}$ MeV and width
$\Gamma=226\pm44\pm87$ MeV \cite{Yuan:2007sj, Liu:2013dau}.
Besides $Y(4260)$, another charmonium-like state $Y(4360)$ was announced by the Babar and Belle collaborations in a similar  process $e^+e^-\to \pi^+\pi^- \psi(3686)$ \cite{Aubert:2007zz, Wang:2007ea, Wang:2014hta}. The spin-parity quantum number of both $Y(4260)$ and $Y(4360)$ produced via the $e^+e^-$ annihilation is  $J^{PC}=1^{--}$   \cite{Aubert:2005rm,Yuan:2007sj,Wang:2007ea}. Thus, theorists first had categorized them into the $\psi$ family. However, this assignment to $Y(4260)$ and $Y(4360)$ is not satisfactory due to the following obstacles: 

(1) There already exist $\psi(4040)$, $\psi(4160)$ and $\psi(4415)$ above 4 GeV in the $\psi$ family \cite{Agashe:2014kda}, so  not much room in the $J/\psi$ family is available for many newly observed charmonium-like states. 

(2) If $Y(4260)$ and $Y(4360)$ are higher excited charmonia, open-charm decays should stand as dominant decay modes contributing to their total widths, however, until now the open-charm decay channels of $Y(4260)$ and $Y(4360)$ are not experimentally observed at all \cite{Abe:2006fj,Pakhlova:2008zza,Pakhlova:2007fq,Pakhlova:2009jv}. Then, the large widths of $Y(4260)$ and $Y(4360)$, i.e., $\Gamma_{Y(4260)}=120\pm 12$ MeV and $\Gamma_{Y(4360)}=78\pm 16$ MeV \cite{Agashe:2014kda}, are hardly understood with that assumption. 

(3) The measurement of $R$ value is an effective approach to identify vector resonances like $\rho$ meson and the charmonia in the $\psi$ family. However, there does not exist any evidence of $Y(4260)$ and $Y(4360)$ in the $R$ value scan \cite{Burmester:1976mn,Brandelik:1978ei,Siegrist:1981zp,Bai:1999pk,CroninHennessy:2008yi,Ablikim:2009ad}. 

(4) The line shapes of the $Y(4260)$ and $Y(4360)$ peaks observed in the two modes
obviously deviate from the Lorentzian form and are obviously asymmetric.

In addition,  $Y(4260)$ and $Y(4360)$ were  proposed to be exotic states, including hybrid charmonium \cite{Zhu:2005hp} and molecular states \cite{Ding:2008gr}. Among these possible exotic structures, the  $D\bar{D}_1$ molecular state \cite{Ding:2008gr} should strongly couple to the nearby $D\bar{D}_1$ final state. However, lack of the signal of $Y(4260)$ in that channel poses a serious challenge to that aforementioned exotic-state explanation  \cite{Liu:2013waa}. Thus we should admit that the identity of
$Y(4260)$ and $Y(4360)$ is not as previously proposed. Due to the existing confusion about $Y(4260)$ and $Y(4360)$, we have re-examined them by trying an alternative approach. The new mechanism under consideration is different from the previous viewpoints and determines that the $Y(4260)$ and $Y(4360)$ peaks are not real resonances. The mechanism indeed offers a reasonable explanation to the exotic behaviors of the observed peaks.
Concretely,  we propose that a unified Fano-like interference would be the responsible scenario.

In physics, the Fano interference which was studied a long time ago for understanding some problems in atomic and nuclear physics, results in a Fano resonance. In that framework, the genuine
mass eigenstate $\phi$ (or $\phi_i$'s) interacts with the continuum via the Fano Hamiltonian and the consequence is that the peak position of $\phi$ is shifted and an additional phase is caused which would distort the Gaussian line shape of the resonance $\phi$ to be asymmetric \cite{fano1,fano2}.

In this work, by checking the experimental data of $Y(4260)$ and $Y(4360)$, we determine that $Y(4260)$ and $Y(4360)$ are not genuine resonances, and the two peaks observed in $e^+e^-\to \pi^+\pi^- J/\psi$ and $e^+e^-\to \pi^+\pi^- \psi(3686)$, actually are consequences of a Fano-like interference of $\psi(4160)$ and $\psi(4415)$ with the continuum contributions. Thus
the line shapes of $Y(4260)$ and $Y(4360)$ observed
in the $e^+e^-\to \pi^+\pi^- J/\psi$ and $e^+e^-\to \pi^+\pi^- \psi(3686)$ are asymmetric \cite{Aubert:2005rm,Yuan:2007sj,Wang:2007ea} due to the extra Fano phase.
Indeed, the two well known charmonia $\psi(4160)$ and $\psi(4415)$ are close to the peak positions of  $Y(4260)$ and $Y(4360)$ \cite{Agashe:2014kda}, and obviously, the Fano
effects could shift them to the positions of $Y(4260)$ and $Y(4360)$. Thus
we are tempted to conclude that
$Y(4260)$ and $Y(4360)$ are not genuine resonances, but the Fano-like interference of $\psi(4160)$ and $\psi(4415)$ with their respective continua results in the  $Y(4260)$ and $Y(4360)$ peaks. This picture
can naturally answer why $Y(4260)$ and $Y(4360)$ are absent in the $R$ value scan \cite{Burmester:1976mn,Brandelik:1978ei,Siegrist:1981zp,Bai:1999pk,CroninHennessy:2008yi,Ablikim:2009ad} and  the measured cross sections of $e^+e^-$ annihilation into open-charm final states do not receive contributions from $Y(4260)$ and $Y(4360)$ at all \cite{Abe:2006fj,Pakhlova:2008zza,Pakhlova:2007fq,Pakhlova:2009jv}.

In atomic physics \cite{fano-atomic}, condensed matter physics \cite{Fano-cm}, and even nuclear physics \cite{Orrigo:2006rd}  the Fano interference phenomena have been
widely studied. Moreover, the interference between the signal of the Higgs resonance in gluon fusion and  the continuum  background for $gg\to\gamma\gamma$ was also considered by Dixon and Li a while ago \cite{Dixon} and it is
a Fano-like effect as a matter of fact.  Along the same lines, a while ago Cao and Lenske \cite{Cao:2014vca} indicated that
the distortion of the $\psi(3770)$ line shape could be understood by the Fano effect. 
Additionally, there were some papers which discussed the importance of this mechanism in $XYZ$ physics \cite{Chen:2010nv,Chen:2011kc,Papinutto:2013uya,Esposito:2014rxa}.   
In this work, we
apply the Fano-like interference picture to explain puzzles for charmonium-like states $Y(4260)$ and $Y(4360)$ and it is a typical example in particle physics at the lower energy
region to discuss hadron spectra.
The success of the application not only reveals the  underlying mechanism resulting in $Y(4260)$ and $Y(4360)$, but also stimulates extensive interest in exploring other Fano interference phenomena in hadronic reactions.


{\it FANO-LIKE INTERFERENCE PICTURE FOR $Y(4260)$ and $Y(4360)$.---}The Fano-like interference refers to the interference between continuum and resonance contributions.  For an analytical calculation on the interference between the continuum and resonance contributions to $e^+e^- \to \pi^+ \pi^- J/\psi$, one needs concrete information about the background.  Since so far, no data on the pure background are available, for describing the continuum contribution  we have to adopt an empirical formula to deal with the background, namely, the  cross section contributed by the continuum background is written as
\begin{eqnarray}
  \includegraphics[width=3.2cm]{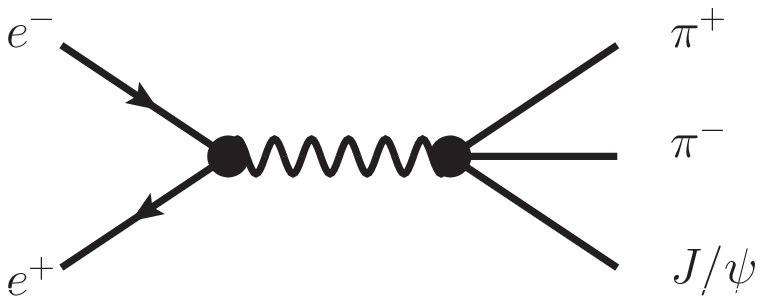}\raisebox{0.5cm}{$\widehat{=} g\, u^2 e^{-a u^2}$} \label{Eq:NonR}
\end{eqnarray}
with $u=\sqrt{s}-\sum_{f} m_f$ being the available kinetic energy, where $\sqrt s$ is total energy in the center-of-mass frame of $e^+e^-$ and
$\sum_{f} m_f$ sums over the masses of all particles in the final state. Additionally, two phenomenological parameters
$a$  and $g$ which are obviously related to non-perturbative QCD, can be treated as free parameter to be determined by fitting the experimental data of $e^+e^- \to \pi^+ \pi^- J/\psi$. The expression indeed gives a smooth curve for the background and this expression is somehow similar to the formula for describing the background in three-body decays of B-mesons, i.e., the Argus function. However since there is no experimental support to the empirical formula, in the following
computations, we will turn to an alternative way to determine the Fano-interference.

Besides the continuum contribution, there exists  contribution  to $e^+e^- \to \pi^+ \pi^- J/\psi$ from the genuine resonances, where
electron and positron annihilate into a virtual photon, which converts into a vector charmonium. Later this charmonium which is on mass shell decays into the final state $\pi^+\pi^- J/\psi$. Then, the mode induced by the intermediate vector charmonium interacts with the direct annihilation of $e^+e^-$ into $\pi^+\pi^- J/\psi$ which stands as the continuum background,via the Fano-like Hamiltonian. Since the Fano-like interaction shifts the peak position of the resonance to the experimentally observed position, thus
the key point is to identify a suitable
intermediate vector charmonium which must possess the required quantum numbers and not be far away from the observed peak. That is the crucial task.
Indeed, we notice that the charmonium-like structure $Y(4260)$ under discussion resides between
two well known higher charmonia $\psi(4160)$ and $\psi(4415)$. We suppose that via a Fano-like interaction, they may produce the $Y(4260)$  and $Y(4360)$ signals.

In general, the
contribution of a higher charmonium to $e^+e^- \to \pi^+ \pi^- J/\psi$   can be described as
\begin{eqnarray}
  \includegraphics[width=3.2cm]{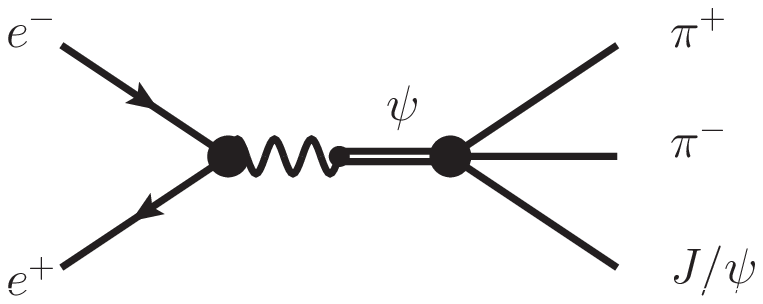}\raisebox{0.5cm}{$\widehat{=} \frac{\sqrt{12\pi \Gamma^{e^+e^-}_{\psi} \times \mathcal{B}(\psi \to \pi^+ \pi^- J/\psi)\Gamma_{\psi}}}{s-m_{\psi}^2+im_{\psi} \Gamma_{\psi} } \sqrt{\frac{\Phi_{\mathrm{2\to3}}(s)}{\Phi_{\mathrm{2\to3}}(m_{\psi}^2)}}$ }, \label{Eq:Res}
\end{eqnarray}
where $\Phi_{2\to 3}(s)$ denotes the phase space for $e^+ e^- \to \pi^+ \pi^- J/\psi$, while $\Phi_{2\to 3}(m_\psi^2)$ is obtained by replacing $s\to m_\psi^2$ in $\Phi_{2\to 3}(s)$. Here $\psi$ is either $\psi(4160)$ or $\psi(4415))$ as the intermediate vector charmonium. The masses and decay widths of $\psi(4160)$ and $\psi(4415)$ take the central values given in the PDG average
data \cite{Agashe:2014kda}. The production rate ${\mathcal R}_\psi\equiv\Gamma^{e^+e^-}_\psi\times \mathcal{B}(\psi \to \pi^+ \pi^- J/\psi)$ is set as a free parameter in our computations and will be determined by fitting data.

After considering the interference between continuum and resonance contributions, the total signal amplitude of $e^+ e^- \to \pi^+ \pi^- J/\psi$ can be parametrized as
\begin{eqnarray}
\mathcal{A}^{\mathrm{Total}}= \mathcal{A}_{\mathrm{Continuum}}  + e^{i \phi_{1}} \mathcal{A}_{\psi(4160)} + e^{i \phi_{2}} \mathcal{A}_{\psi(4415)},\label{Eq:total}
\end{eqnarray}
where $\mathcal{A}_{\mathrm{Continuum}}$, $\mathcal{A}_{\psi(4160)}$ and $\mathcal{A}_{\psi(4415)}$ are defined in Eqs. (\ref{Eq:NonR})-(\ref{Eq:Res}). In this signal amplitude, there exist 6 free parameters, which are,
\begin{eqnarray}
&&g,\ a,\ \mathcal{R}_{\psi(4160)},\  \mathcal{R}_{\psi(4415)}, \ \phi_1, \ \phi_2. \nonumber
\end{eqnarray}

\begin{figure}[htbp]
  \centering
  \includegraphics[width=8cm]{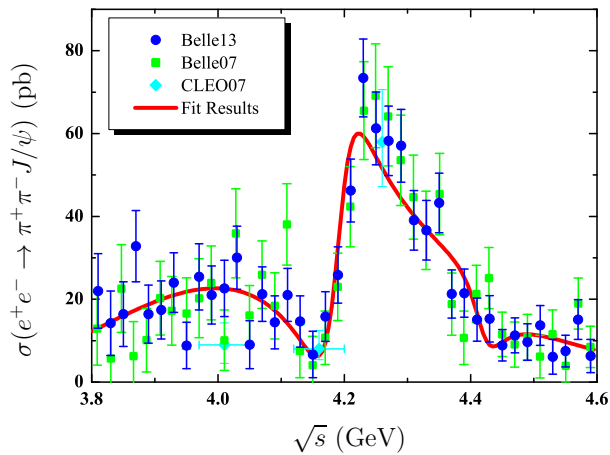}
  \caption{(color online). The obtained result (red curve) of the cross section for $e^+ e^- \to \pi^+ \pi^- J/\psi$ in the Fano-like interference picture and a comparison with the experimental data of the cross section for $e^+ e^- \to \pi^+ \pi^- J/\psi$ \cite{Yuan:2007sj, Liu:2013dau,Coan:2006rv} is provided.}\label{Fig:y4260}
\end{figure}

With this constructed simple and explicit model, we analyze the experimental data of the cross sections for $e^+ e^- \to \pi^+ \pi^- J/\psi$ \cite{Liu:2013dau} for testing whether the charmonium-like structure $Y(4260)$ can be reproduced via the Fano-like interference. The fitted the cross section of $e^+ e^- \to \pi^+ \pi^- J/\psi$ is shown in Fig.\ \ref{Fig:y4260}, where the experimental data are taken from publications of the Belle  \cite{Yuan:2007sj, Liu:2013dau} and CLEO collaborations \cite{Coan:2006rv}.

From Fig.\ \ref{Fig:y4260}, one may conclude: (1) the asymmetric $Y(4260)$ line shape can be well reproduced by the Fano-like interference, which provides direct evidence that the observed $Y(4260)$ is a fake resonance. (2) Just as indicated in Ref. \cite{Yuan:2007sj},
there exists a broad structure, which corresponds to $Y(4008)$. Our result shows that this broad structure $Y(4008)$, which is a companion peak to $Y(4260)$, is also induced by the Fano-like interference. It implies that the charmonium-like structure $Y(4008)$ is also not a genuine resonance.

\renewcommand{\arraystretch}{1.5}
\begin{table}[htbp]
  \centering
\caption{The parameters obtained by fitting the experimental data of the cross sections for $e^+ e^- \to \pi^+ \pi^- J/\psi$ \cite{Liu:2013dau} and $e^+e^-\to \pi^+\pi^- \psi(3686)$ \cite{Wang:2014hta}.} \label{Tab:y4260}
\begin{tabular}{cccc}
  \toprule[1pt] \toprule[1pt]
\multicolumn{4}{c}{$e^+e^-\to \pi^+\pi^- J/\psi$}\\
\midrule[1pt]
$g$   &     $(49.1 \pm  3.6)\ \mathrm{GeV}^{-1}$ &
$a$                &   $(1.9  \pm  0.1)\ \mathrm{GeV}^{-2}$ \\
$\mathcal{R}_{\psi(4160)}$            &   $(2.6  \pm  0.7)\ \mathrm{eV}$ &
$\phi_1$             &   $(6.0  \pm  0.1)\ \mathrm{rad}$ \\
$\mathcal{R}_{\psi(4415)}$           &   $(5.2  \pm  0.7) \ \mathrm{eV}$ &
$\phi_2$             &   $(4.4  \pm  0.1)\ \mathrm{rad}$ \\
\midrule[1pt]
\multicolumn{4}{c}{$e^+e^-\to \pi^+\pi^- \psi(3686)$}\\
\midrule[1pt]
$g$   &     $(150.6   \pm     9.8)\ \mathrm{GeV}^{-1} $ &
$a$                &         $(5.6     \pm     0.7)\ \mathrm{GeV}^{-2} $ \\
$\mathcal{R}_{\psi(4160)}$   &   $(1.5     \pm     0.8)\ \mathrm{eV} $ &
$\phi_1$             &             $(4.2     \pm     0.2)\ \mathrm{rad} $ \\
$\mathcal{R}_{\psi(4415)}$   &  $(2.3     \pm     0.8) \ \mathrm{eV}$ &
$\phi_2$             &                   $(3.6     \pm     0.2)\ \mathrm{rad} $\\
\bottomrule[1pt]
  \bottomrule[1pt]
\end{tabular}
\end{table}

The parameters determined by fitting the experimental data of the cross section of $e^+ e^- \to \pi^+ \pi^- J/\psi$ are listed in Table \ref{Tab:y4260}, which is believed to provide valuable information toward further studying similar effects.

Now we would like to investigate what other consequences and predictions we may draw from this scenario.

The production rates $\mathcal{R}_{\psi(4160)}\equiv\Gamma_{\psi(4160)}^{e^+ e^-}\mathcal{B}(\psi(4160) \to  \pi^+ \pi^- J/\psi)$ and $\mathcal{R}_{\psi(4415)}\equiv\Gamma_{\psi(4415)}^{e^+ e^-}\mathcal{B}(\psi(4415) \to \pi^+ \pi^- J/\psi )$ are fitted to be $(2.6 \pm 0.7)\ \mathrm{eV}$ and $(5.2 \pm 0.5)\ \mathrm{eV}$, respectively. By fitting the data of $e^+e^-\to \pi^+\pi^- J/\psi$ (both line shape and cross section), we gain
the production rates $\mathcal{R}_{\psi(4160)}$ and $\mathcal{R}_{\psi(4415)}$ which are products of two factors. Once $\Gamma_{\psi(4160)}^{e^+ e^-}$ is known, we would immediately
calculate the branching ratio $\mathcal{B}(\psi(4160) \to \pi^+ \pi^- J/\psi)$ which can be compared with the directly measured data of the three-body decay. If the two values are
consistent, our scenario would be further confirmed.

Using the dilepton partial widths $\Gamma_{\psi(4160)}^{e^+ e^-}= (0.48 \pm 0.22) \,\mathrm{keV}$ and $\Gamma_{\psi(4415)}^{e^+ e^-}=(0.58 \pm 0.07)\  \,\mathrm{keV}$ \cite{Agashe:2014kda}, we get the branching ratios of $\psi(4160) \to \pi^+ \pi^- J/\psi$ and $\psi(4415) \to \pi^+ \pi^- J/\psi$ to be $(5.4 \pm 3.9) \times 10^{-3}$ and $(8.9 \pm 2.3)\times 10^{-3}$, respectively. Unfortunately, however, so far the experimental measurements on the branching ratios of such modes do not reach a satisfactory accuracy  level, i.e, the corresponding errors are across a wide range, so that any solid conclusion are hard to make at present yet. However, on the optimistic aspect, the rough measurements still provide valuable information for theoretical studies. The upper limit of  $\mathcal{B}(\psi(4160) \to \pi^+ \pi^- J/\psi )$ was measured to be $3 \times 10^{-3}$ by the CLEO Collaboration \cite{Coan:2006rv}, which is quoted by Particle Data Group \cite{Agashe:2014kda}. Our results are consistent with the available data even though they are not precise.

\begin{figure}[htpb]
  \centering
  \includegraphics[width=8cm]{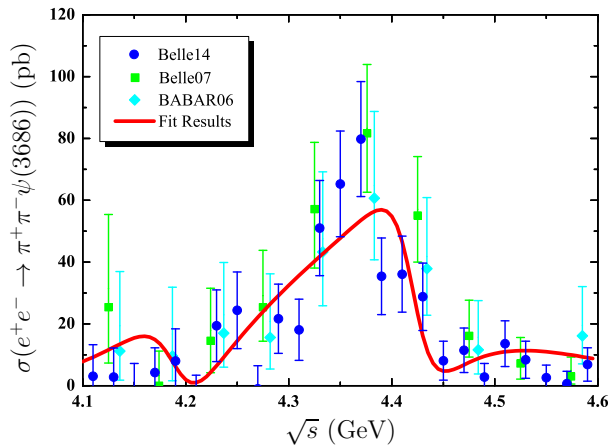}
  \caption{(color online). The same as Fig.\ \ref{Fig:y4260} but for the line shape of the cross section for $e^+ e^- \to \pi^+ \pi^- \psi(2S)$. The experimental data are taken from the measurement of the BaBar \cite{Aubert:2007zz} and the Belle \cite{Wang:2014hta, Wang:2007ea} collaborations. }\label{Fig:y4360}
\end{figure}

When the $Y(4260)$ signal is successfully reproduced through the Fano-like interference picture, we naturally extend the same mechanism to study another charmonium-like structure $Y(4360)$ observed in $e^+e^- \to \pi^+ \pi^-\psi(3686)$.
Our model in Eq.\ (\ref{Eq:total}) also applies to this investigation. By fitting the experimental data of the cross sections of $e^+e^- \to \pi^+ \pi^-\psi(3686)$ \cite{Wang:2014hta}, we obtain the values of fitted parameters which are listed in the lower part of the Table \ref{Tab:y4260} altogether with that for $e^+e^- \to \pi^+ \pi^-J/\psi$.
The corresponding energy distribution curve of the   $e^+e^- \to \pi^+ \pi^-\psi(3686)$ cross section which is calculated in our model is shown in Fig.\ \ref{Fig:y4360} meanwhile the experimental data  \cite{Wang:2014hta, Wang:2007ea, Aubert:2007zz} are also presented in the same figure for comparison.
Similar to the case of $Y(4260)$ observed in the cross sections of $e^+ e^- \to \pi^+ \pi^- J/\psi$, the structure $Y(4360)$ is also asymmetric. Here, introducing the Fano-like interference between the continuum and $\psi(4160)/\psi(4415)$ resonance contributions, the typical  asymmetric structure around 4.36 GeV appearing in the $e^+e^- \to \pi^+ \pi^-\psi(3686)$  energy distribution is well reproduced. It is noted that unlike $e^+e^- \to \pi^+ \pi^-J/\psi$, at the energy distribution of $e^+e^- \to \pi^+ \pi^-\psi(3686)$, no extra peak
shows up, thus we conclude that appearance of the $Y(4008)$  peak is due to the interference of $\psi(4160)$/$\psi(4415)$ with the continuum of $e^+e^- \to \pi^+ \pi^-J/\psi$
only.

The fitted results determine the production rates $\Gamma_{\psi(4160)}^{e^+ e^-}\mathcal{B}(\psi(4160) \to  \pi^+ \pi^- \psi(2S))$ and $\Gamma_{\psi(4415)}^{e^+ e^-}\mathcal{B}(\psi(4415) \to  \pi^+ \pi^- \psi(2S))$ to be $(1.5 \pm 0.8)\ \mathrm{eV}$ and $(2.3 \pm 0.8)\ \mathrm{eV}$. Then, the branching ratios of $\psi(4160) \to \pi^+ \pi^- \psi(3686)$ and $\psi(4415) \to \pi^+ \pi^- \psi(3686)$ can be estimated as $(3.1 \pm 3.0) \times 10^{-3}$ and $(4.0 \pm 1.9) \times 10^{-3}$ respectively.  The upper bound of $\mathcal{B}(\psi(4160) \to \pi^+ \pi^- J/\psi)$ was reported to be $4\times 10^{-3}$ by the CLEO Collaboration \cite{Coan:2006rv}, which is also consistent with our results.
Since those branching ratios have not been well experimentally measured so far, we suggest to carry out further experimental study on these hidden-charm dipion decays  up to a high accuracy.

\begin{figure}[htbp]
  \centering
  \includegraphics[width=4.2cm]{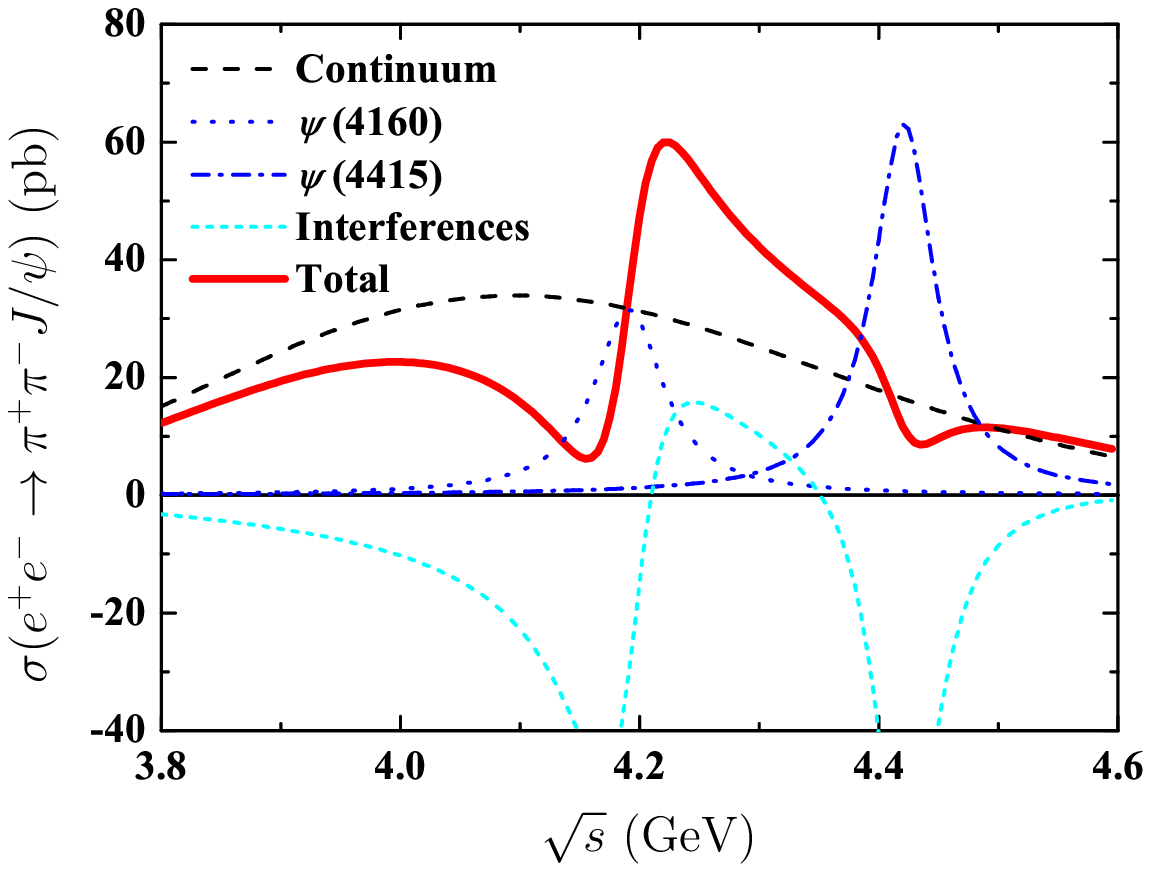}
  \includegraphics[width=4.22cm]{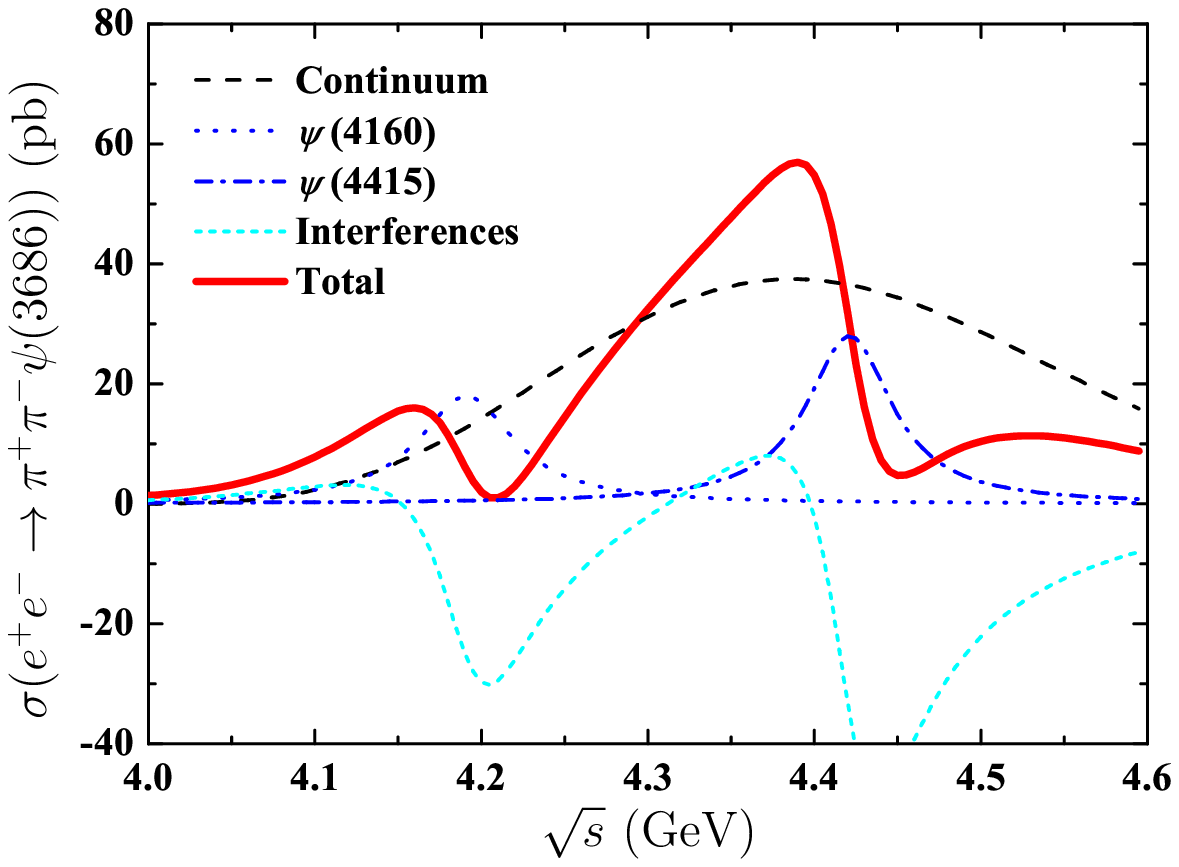}

  \caption{(color online). The separate contributions from the background, resonances and the interferences.}\label{Fig:sep}
\end{figure}

To understand the functions of the resonances and the continuum background, in Fig.\ \ref{Fig:sep}, their individual contributions to the mass spectra are shown along with  the final result where the interferences are accounted for. It is interesting to notice that the line shapes of the interferences in the cross sections for $e^+ e^- \to \pi^+ \pi^- J/\psi$ and  $e^+ e^- \to \pi^+ \pi^- \psi(2S)$ are very similar. In addition, we also check the obtained fitted parameters listed in Table \ref{Tab:y4260}, and find  similarity in the corresponding fitted parameters, which reflects the similarity between $e^+e^- \to \pi^+ \pi^-\psi(3686)$ and $e^+ e^- \to \pi^+ \pi^- J/\psi$.

{\it CONCLUSION.---}So far, there were five charmoinum-like states $Y(4260)$, $Y(4360)$, $Y(4008)$, $Y(4630)$ and $Y(4660)$ observed through the
$e^+e^-$ annihilation \cite{Liu:2013waa}. The corresponding hidden-charm dipion productions may play an important role for studying those charmonium-like states. In this paper, we analyze the asymmetric line shapes of $Y(4260)$ and $Y(4360)$ and propose a unified Fano-like interference picture, by which
the $Y(4260)$, $Y(4360)$, and $Y(4008)$ structures are well reproduced. Indeed, $Y(4260)$, $Y(4360)$, $Y(4008)$ are all not genuine resonances. The Fano-like interference reasonably answers why $Y(4260)$, $Y(4360)$, and $Y(4008)$ signals are absent in the $R$ value scan and the corresponding open-charm decay modes have never been observed. All these puzzles have been existing in studies on these charmonium-like states for a long time since $Y(4260)$, $Y(4360)$, $Y(4008)$ were reported by several experimental collaborations.

For higher charmonia, their open-charm decays may compose the dominant modes, but meanwhile their hidden-charm dipion (or dikaon)  decays are also non-negligible. Unfortunately, so far, measurements on the two crucial modes are only at preliminary stages, i.e., no reliable data are available.
Thus, from the theoretical aspect, we believe that two well-established higher charmonia $\psi(4160)$ and $\psi(4415)$ should decay into $\pi^+\pi^-J/\psi$ and $\pi^+\pi^-\psi(3686)$
with relatively large branching ratios. However,
when checking the experimental data of $e^+e^-\to \pi^+\pi^-J/\psi$ \cite{Aubert:2005rm,Yuan:2007sj,Liu:2013dau} and $e^+e^-\to \pi^+\pi^-\psi(3686)$ \cite{Aubert:2007zz,Wang:2007ea,Wang:2014hta}, we notice absence of the $\psi(4160)$ and $\psi(4415)$ signals. Alternately, explicit $Y(4260)$ and $Y(4360)$ structures exist in the corresponding $\pi^+\pi^-J/\psi$ and $\pi^+\pi^-\psi(3686)$ distributions. By the Fano-like interference picture, higher charmonia $\psi(4160)$ and $\psi(4415)$ interfere with the continuum contribution to result in the $Y(4260)$ and $Y(4360)$ structures. Finally, $\psi(4160)$ and $\psi(4415)$ are not the typical Breit-Wigner distribution to be responsible for these measured $e^+e^-\to \pi^+\pi^-J/\psi$ \cite{Aubert:2005rm,Yuan:2007sj,Liu:2013dau} and $e^+e^-\to \pi^+\pi^-\psi(3686)$ cross sections. By our model, we obtain the branching ratios of  $\psi(4160)$ and $\psi(4415)$ decaying into $\pi^+\pi^-J/\psi$ and $\pi^+\pi^-\psi(3686)$. The results will be checked by future precise experimental measurements, so that they are not only
valuable information for the studies on $\psi(4160)$ and $\psi(4415)$, but also can be applied to verify the proposed picture.

In summary, the present study provides an optimistic scenario to understand these peculiar charmonium-like states. It is believed that the proposed mechanism may stimulate more extensive discussions about  charmonium-like states. Before closing this work, we would like to persuade our experimental colleagues to carry out a study on similar processes $e^+e^-\to \pi^+\pi^-\psi(3770)$ and $e^+e^-\to K^+K^- J/\psi$, where  the resonance contributions of $\psi(4160)$ and $\psi(4415)$ would interfere with the continuum contribution. Such  Fano-like interference phenomena should be observed by
BESIII and forthcoming BelleII which have a remarkable opportunity to make solid confirmation about the mechanism (or negate it).

\section*{Acknowledgments}

This project is supported by the National Natural Science Foundation
of China under Grants No. 11222547, No. 11175073, No. 11375240, No. 11375128 and No. 11135009.


\begin{thebibliography}{99}
\bibitem{Liu:2013waa}
  X.~Liu,
  An overview of $XYZ$ new particles,
  Chin.\ Sci.\ Bull.\  {\bf 59}, 3815 (2014).

\bibitem{Aubert:2005rm}
  B.~Aubert {\it et al.} [BaBar Collaboration],
  Observation of a broad structure in the $\pi^+ \pi^- J/\psi$ mass spectrum around 4.26 GeV/c$^2$,
  Phys.\ Rev.\ Lett.\  {\bf 95}, 142001 (2005).

\bibitem{Yuan:2007sj}
  C.~Z.~Yuan {\it et al.} [Belle Collaboration],
  Measurement of $e^+ e^- \to \pi^+ \pi^- J/\psi$ cross-section via initial state radiation at Belle,
  Phys.\ Rev.\ Lett.\  {\bf 99}, 182004 (2007).

\bibitem{Liu:2013dau}
  Z.~Q.~Liu {\it et al.} [Belle Collaboration],
 Study of $e^+e^- \to \pi^+\pi^- J/\psi$ and Observation of a Charged Charmoniumlike State at Belle,
  Phys.\ Rev.\ Lett.\  {\bf 110}, 252002 (2013).

\bibitem{Aubert:2007zz}
  B.~Aubert {\it et al.} [BaBar Collaboration],
  Evidence of a broad structure at an invariant mass of 4.32 $\mathrm{GeV}/c^{2}$ in the reaction $e^{+} e^{-} \to \pi^{+} \pi^{-} \psi{({2S})}$ measured at BaBar,
  Phys.\ Rev.\ Lett.\  {\bf 98}, 212001 (2007).


\bibitem{Wang:2007ea}
  X.~L.~Wang {\it et al.} [Belle Collaboration],
  Observation of Two Resonant Structures in $e^+ e^- \to \pi^+ \pi^- \psi(2S)$ via Initial State Radiation at Belle,
  Phys.\ Rev.\ Lett.\  {\bf 99}, 142002 (2007).

\bibitem{Wang:2014hta}
  X.~L.~Wang {\it et al.} [Belle Collaboration],
  Measurement of $e^+e^- \to \pi^+\pi^-\psi(2S)$ via Initial State Radiation at Belle,
  Phys.\ Rev.\ D {\bf 91},  112007 (2015).


\bibitem{Agashe:2014kda}
  K.~A.~Olive {\it et al.} [Particle Data Group Collaboration],
 Review of Particle Physics,
  Chin.\ Phys.\ C {\bf 38}, 090001 (2014).

\bibitem{Abe:2006fj}
  K.~Abe {\it et al.} [Belle Collaboration],
Measurement of the near-threshold $e^+ e^- \to D^{*\pm} D^{*\mp}$ cross section using initial-state radiation,
  Phys.\ Rev.\ Lett.\  {\bf 98}, 092001 (2007).

\bibitem{Pakhlova:2008zza}
  G.~Pakhlova {\it et al.} [Belle Collaboration],
  Measurement of the near-threshold $e^+ e^- \to D \bar D$ cross section using initial-state radiation,
  Phys.\ Rev.\ D {\bf 77}, 011103 (2008).

\bibitem{Pakhlova:2007fq}
  G.~Pakhlova {\it et al.} [Belle Collaboration],
  Observation of $\psi(4415)\to D \bar{D}^*_2(2460)$ decay using initial-state radiation,
  Phys.\ Rev.\ Lett.\  {\bf 100}, 062001 (2008).

\bibitem{Pakhlova:2009jv}
  G.~Pakhlova {\it et al.} [Belle Collaboration],
 Measurement of the $e^+ e^- \to D^0 D^{*-} \pi^+$ cross section using initial-state radiation,
  Phys.\ Rev.\ D {\bf 80}, 091101 (2009).


\bibitem{Burmester:1976mn}
  J.~Burmester {\it et al.} [PLUTO Collaboration],
  The Total Hadronic Cross-Section for $e^+ e^-$ Annihilation Between 3.1 GeV and 4.8 GeV Center-Of-Mass Energy,
  Phys.\ Lett.\ B {\bf 66}, 395 (1977).

\bibitem{Brandelik:1978ei}
  R.~Brandelik {\it et al.} [DASP Collaboration],
  Total Cross-section for Hadron Production by $e^+ e^-$ Annihilation at Center-of-mass Energies Between 3.6 {GeV} and 5.2 {GeV},
  Phys.\ Lett.\ B {\bf 76}, 361 (1978).

\bibitem{Siegrist:1981zp}
  J.~Siegrist {\it et al.},
Hadron Production by $e^+ e^-$ Annihilation at Center-Of-Mass Energies Between 2.6 GeV and 7.8 GeV. Part 1. Total Cross-Section, Multiplicities and Inclusive Momentum Distributions,
  Phys.\ Rev.\ D {\bf 26}, 969 (1982).

\bibitem{Bai:1999pk}
  J.~Z.~Bai {\it et al.} [BES Collaboration],
  Measurement of the total cross-section for hadronic production by $e^+ e^-$ annihilation at energies between 2.6 GeV-5 GeV,
  Phys.\ Rev.\ Lett.\  {\bf 84}, 594 (2000).

\bibitem{CroninHennessy:2008yi}
  D.~Cronin-Hennessy {\it et al.} [CLEO Collaboration],
  Measurement of Charm Production Cross Sections in $e^+e^-$ Annihilation at Energies between 3.97 and 4.26 GeV,
  Phys.\ Rev.\ D {\bf 80}, 072001 (2009).

\bibitem{Ablikim:2009ad}
  M.~Ablikim {\it et al.} [BES Collaboration],
  $R$ value measurements for $e^+ e^-$ annihilation at 2.60 GeV, 3.07 GeV and 3.65 GeV,
  Phys.\ Lett.\ B {\bf 677}, 239 (2009).

\bibitem{Zhu:2005hp}
  S.~L.~Zhu,
  The Possible interpretations of $Y(4260)$,
  Phys.\ Lett.\ B {\bf 625}, 212 (2005).

\bibitem{Ding:2008gr}
  G.~J.~Ding,
  Are $Y(4260)$ and $Z^+_2(4250)$ are $D_1 D$ or $D_0 D^*$ Hadronic Molecules?,
  Phys.\ Rev.\ D {\bf 79}, 014001 (2009).

\bibitem{fano1}U.~Fano, Effects of Configuration Interaction on Intensities and Phase Shifts, Phys. Rev. {\bf 124}, 1866 (1961).

\bibitem{fano2}A.~Bianconi, Ugo Fano and shape resonances, AIP Conf. Proc. {\bf 652}, 13 (2003).

\bibitem{fano-atomic}C.~Ott, A.~Kaldun, P.~Raith, K.~Meyer, M.~Laux, J.~Evers, C.~H.~Keitel, C.~H.~Greene, T.~Pfeifer, Fano in Spectral Line Shapes: A Universal Phase and Its Laser Control, Science {\bf 340}, 716 (2013).

\bibitem{Fano-cm}H.~G. Luo, T.~Xiang, X.~Q.~Wang, Z.~B.~Su, and L.~Yu, Fano Resonance for Anderson Impurity Systems,
Phys. Rev. Lett. {\bf 92}, 256602 (2004).

\bibitem{Dixon}L.~J.~Dixon and Y.~Li,
  Bounding the Higgs Boson Width Through Interferometry,
  Phys.\ Rev.\ Lett.\  {\bf 111}, 111802 (2013).


\bibitem{Orrigo:2006rd}
  S.~E.~A.~Orrigo, H.~Lenske, F.~Cappuzzello, A.~Cunsolo, A.~Foti, A.~Lazzaro, C.~Nociforo, and J.~S.~Winfield,
  Core excited Fano-resonances in exotic nuclei,
  Phys.\ Lett.\ B {\bf 633}, 469 (2006).

\bibitem{Cao:2014vca}
  X.~Cao and H.~Lenske,
  Charmonium resonances and Fano line shapes,
  arXiv:1408.5600 [nucl-th].

\bibitem{Chen:2010nv} 
  D.~Y.~Chen, J.~He and X.~Liu,
  ``Nonresonant explanation for the Y(4260) structure observed in the $e^+e^-\to J/\psi\pi^+\pi^-$ process,''
  Phys.\ Rev.\ D {\bf 83}, 054021 (2011)
  [arXiv:1012.5362 [hep-ph]].

\bibitem{Chen:2011kc} 
  D.~Y.~Chen, J.~He and X.~Liu,
  ``A Novel explanation of charmonium-like structure in $e^+e^-\to \psi(2S)\pi^+\pi^-$,''
  Phys.\ Rev.\ D {\bf 83}, 074012 (2011)
  [arXiv:1101.2474 [hep-ph]].
    
\bibitem{Papinutto:2013uya} 
  M.~Papinutto, F.~Piccinini, A.~Pilloni, A.~D.~Polosa and N.~Tantalo,
  ``A Tentative Description of $Z_{c,b}$ States in Terms of Metastable Feshbach Resonances,''
  arXiv:1311.7374 [hep-ph].
  
\bibitem{Esposito:2014rxa} 
  A.~Esposito, A.~L.~Guerrieri, F.~Piccinini, A.~Pilloni and A.~D.~Polosa,
  ``Four-Quark Hadrons: an Updated Review,''
  Int.\ J.\ Mod.\ Phys.\ A {\bf 30}, 1530002 (2015)
  [arXiv:1411.5997 [hep-ph]].
    


\bibitem{Coan:2006rv}
  T.~E.~Coan {\it et al.} [CLEO Collaboration],
 Charmonium decays of $Y(4260)$, $\psi(4160)$ and $\psi(4040)$,
  Phys.\ Rev.\ Lett.\  {\bf 96}, 162003 (2006).




\end{thebibliography}
\end{document}